\documentclass[12pt, onecolumn, a4paper]{article}

\usepackage{amssymb}
\usepackage{authblk}
\usepackage[pdftex]{graphicx}
\usepackage{geometry}
\usepackage{cite}

\title{Emergence of long-term rhythmicity within a frustrated triangle oscillator-network}
\author[a]{Masatomo Matsushima\thanks{Corresponding author ({\it Email address}: {\tt mamatsus@mail.doshisha.ac.jp})}}
\author[a]{Hiroshi Ueno}
\author[a]{Yoshiki Kamiya}
\author[b]{Hiroshi Kawakami}

\affil[a]{Faculty of Life and Medical Sciences, Doshisha University, Kyoto, 610-0394, Japan}
\affil[b]{Faculty of Engineering, Tokushima University, Tokushima 770-8506, Japan}

\date{}

\begin{document}
\maketitle

\begin{abstract}
This study tries  to simulate a brain cell network using an electric circuit oscillator called 	“electronic firefly".
Multiple stability was observed in the electric circuit oscillator which is expressed by simple mathematical models.

 \vspace{\baselineskip}
\end{abstract}

\section{Introduction}

The human cerebrum is a very complex tissue consisting of over 10 billion neurons.
The structure of neural circuits in the cerebral cortex has been studied for over 100 years.
However, the complexity of the cerebrum has many unknown points and this is a major obstacle to understanding brain functions. 
Above all, it was unclear whether there was a structure in which a single neural circuit repeated in the cerebral cortex.
Maruoka\cite{Maruoka2011, Maruoka2017} found that in the mouse brain, 
subcortical projection cells formed elongated clusters among cells, 
and their structures were repeatedly arranged in a honeycomb-like hexagonal lattice arrangement. 
In addition, it has been clarified that the neurons included in this cluster receive input from the same neuron, 
indicating that the input may bring about synchronized neural activity.
There are many studies on the synchronization of oscillators, with neurons as oscillators\cite{Sebastian2018, Dion2018, Vodenicarevic2018, Popescu2018, Nikonov2015, Csaba2018, Ziegler2018, Brunner2018, Jeong2018, Coulombe2017}.

A large number of researched neural networks (artificial neural networks) in modern times are controlled by the clock frequency of the CPU.
Of couse, it is well known that the mechanism of the actual brain cell network is different\cite{sulymenko2018}.
Therefore, we are aiming to construct a neural network that incorporates the concept of time control, which may be a network of brain cells.
For that purpose, each cell is replaced by an electric circuit oscillator with a light emitting element and a light receiving element.
A network of multiple electrical circuit oscillators can theoretically be used to simulate brain cell networks\cite{Oumate2018}.

In this study, we introduce the synchronization phenomena caused by three electric circuit oscillators.
The electric circuit oscillator used is well known as “electronic firefly"\cite{Kousaka1998}.
The state of synchronization of the electronic fireflies is in-phase synchronization in which all the oscillators emit and turn off repeatedly and occur at the same timing.
In this synchronization, there is only one aspect of synchronization.
The circuit of this electronic firefly can cause the synchronization of the anti-phase by slightly changing the configuration.
The anti-phase synchronization has multiple stable synchronization\cite{Wu2012, Astakhov2016, Zhichun2016, Aonishi1999}.
In this study, we show multistability from experiments by using an electric circuit oscillator that causes this anti-phase synchronization.
Moreover, we use the well known “Kuramoto model” to discuss the behavior of those synchronization phenomena and show the possibility of simulating brain cell networks.

\section{Hypothesis}

\begin{figure}[h]
\begin{center}
		\includegraphics[width=7cm]{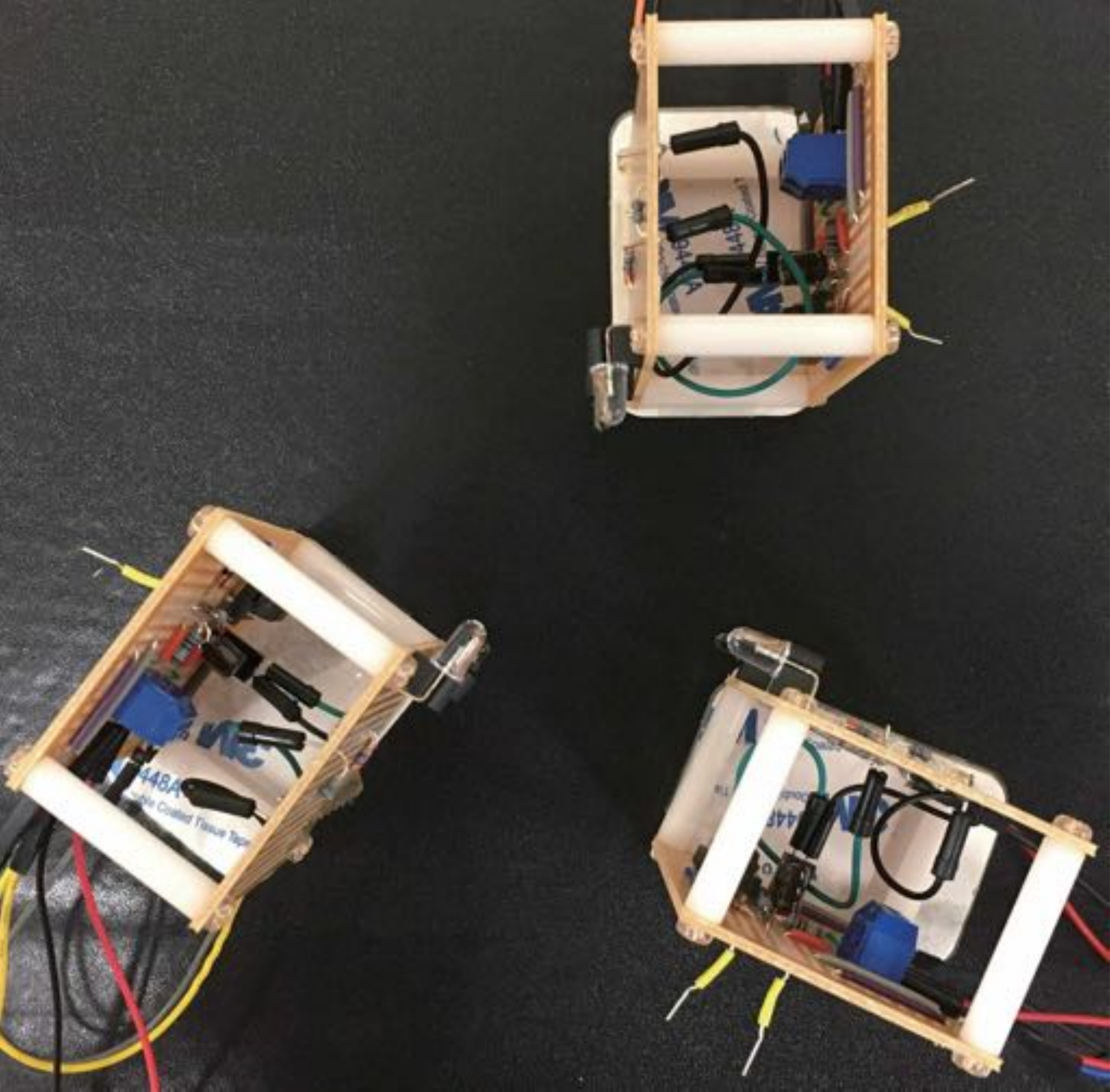}
		\caption{Circuit arrangement example}
		\label{fig:3-os}
\end{center}
\end{figure}

Subcortical projection cells are 3D in a hexagonal lattice arrangement.
It is difficult to model the network between those cells as it is.
So we focused on the fact that a hexagonal lattice array is a collection of triangular lattice arrays.
In other words, we consider it as a network of triangular grids in a 2D model\cite{Giver2011, Levnajic2011, Kralemann2011, Vanag2019, Esfahani2016, Liu2018}.
In this study, we focus on a single triangular lattice.
An electrical circuit oscillator with a light emitting element and a light receiving element is placed at each vertex of the triangular lattice.
An example of the arrangement is shown in Fig.\ref{fig:3-os}.

These oscillators change the behavior of vibration by receiving a stimulus.
Therefore, due to the arrangement of the oscillators, the strength of the stimulus received by the oscillators changes, and it is possible to see many synchronized patterns of light.

In the next section, we explain the circuit used for the oscillator.

\section{Experimental system}

\begin{figure}[h]
\begin{center}
		\includegraphics[width=15cm]{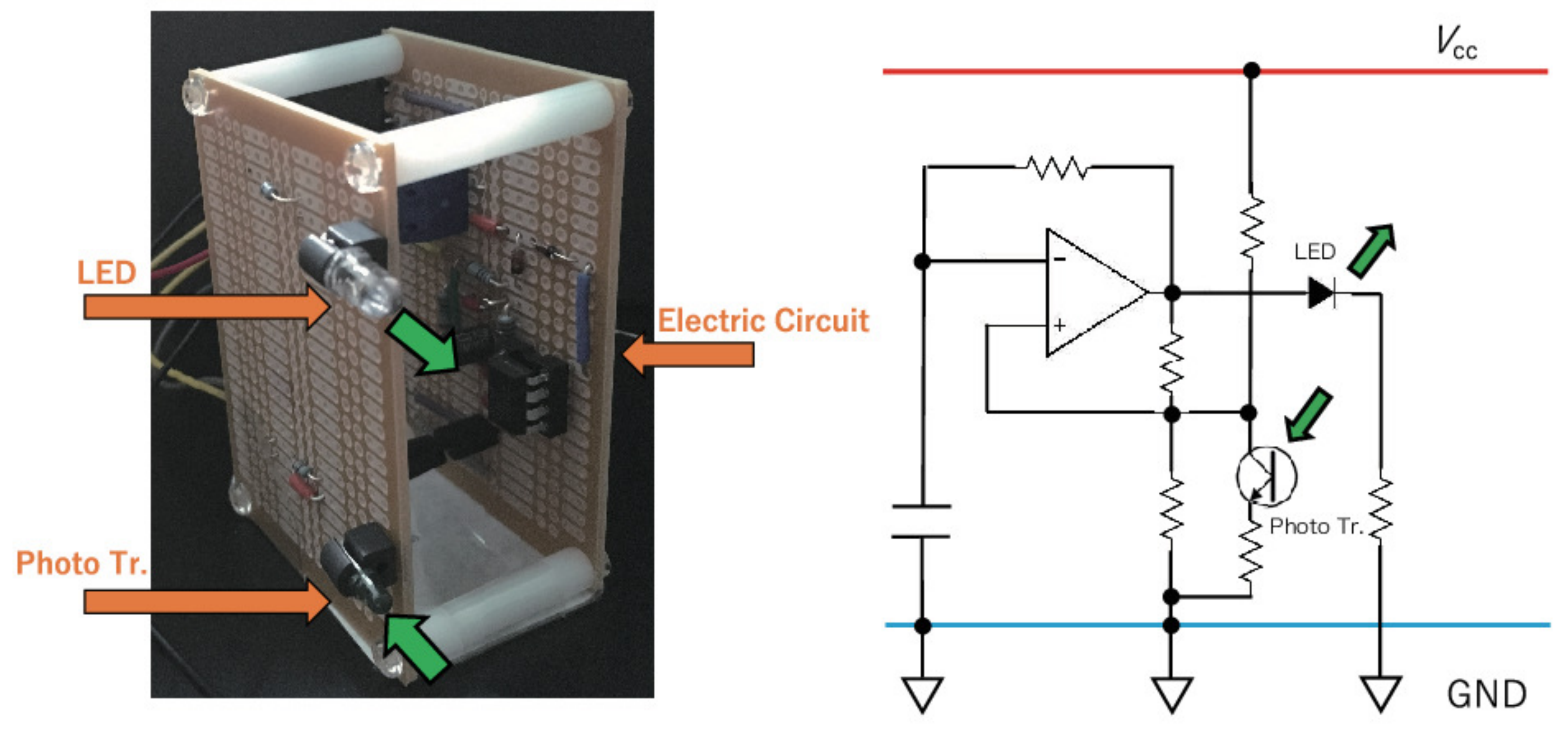}
		\caption{Actual oscillator (left), Circuit Diagram (right)}
		\label{fig:o-os}
\end{center}
\end{figure}

We introduce the electric circuit oscillator used in this experiment.
Fig.\ref{fig:o-os} shows a circuit diagram of the oscillator used.

This electrical oscillator is a well known square wave oscillator circuit and is called an astable multivibrator.

The circuit used in the experiment incorporates a photo transistor as a light emitting element and an LED as a light receiving element.
These electrical oscillators can see the blinking of the LED automatically if they do not receive outside stimulation.
However, when a photo transistor receives light, the value of the resistance in the circuit changes due to the intensity, and charge / discharge characteristics of the capacitor.
As a result, the output time changes, causing a change in the blinking duration and frequency of the LED.
The circuit configuration is almost the same as the well known “electronic firefly” circuit that causes in-phase synchronization.
The difference is that the connection of a photo transistor is reversed.
This causes the synchronization to occur anti-phase.

Next, layout of the electric circuit oscillator is explained again.
The oscillators are all the same circuit design.
The power supply is a common one so that the power can be input simultaneously.
As a matter of course, rectification is performed to prevent coupling synchronization by electric wires.
The LED, which is a light emitting element, uses a green LED for all three oscillators.
In this experiment, the power supply was set to 5V and 300mA, and the capacitor was always discharged before the power was turned on.

The next section shows some interesting results from this experiment.

\section{Results}

\begin{figure}[h]
	\begin{center}
		\includegraphics[width=8cm]{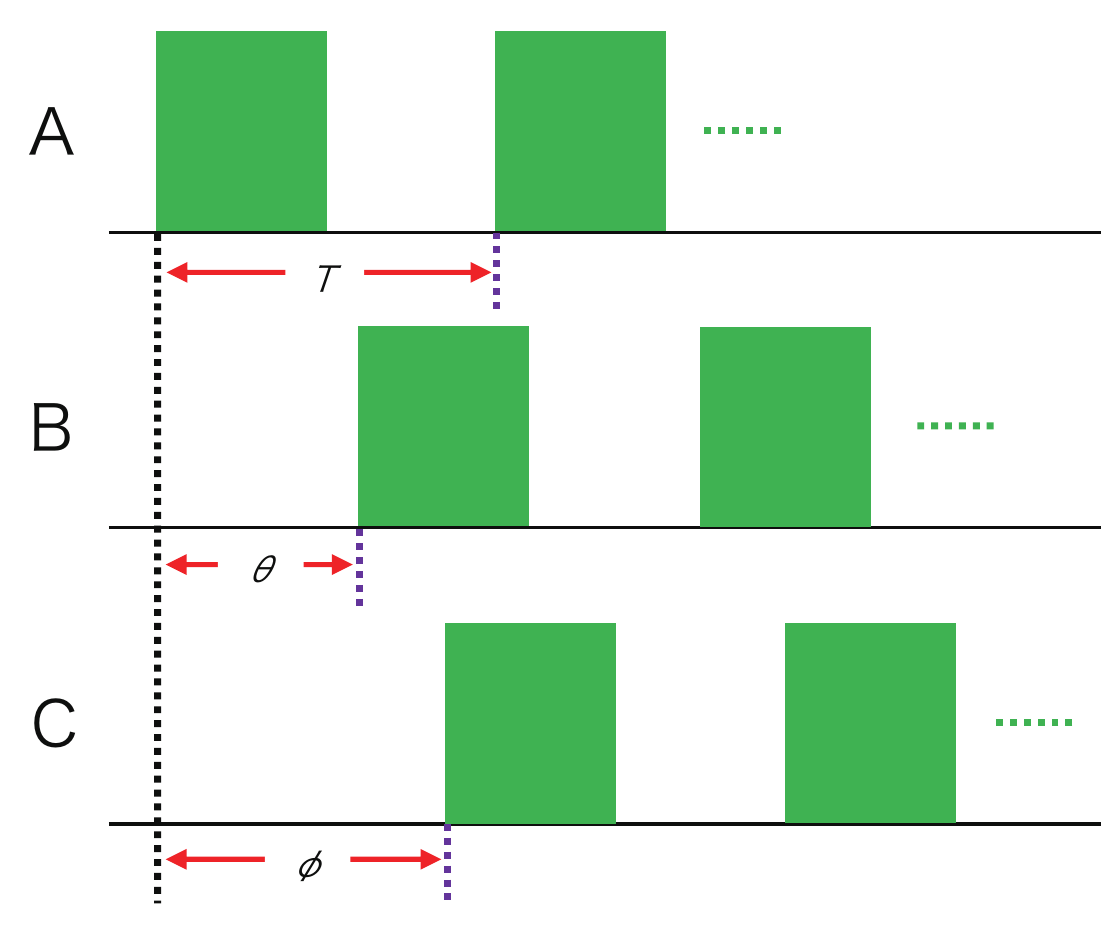}
		\caption{Calculation of phase difference}
		\label{fig:cpd}
	\end{center}
\end{figure}

First, before showing the experimental results. 
We specify the analysis method.
In this study, we focus on the phase difference of the oscillators\cite{Farcot2019, Buskermolen2019, Goldstein2015, Matheny2019}.
We explain with Fig.\ref{fig:cpd}.

Let three oscillators be A, B, and C respectively.
When the light emission start time of the oscillator A is 0, the time until the next light emission is represented by $T$ as one cycle.
Let $\theta$ be the difference between the light emission start of the oscillator B and the light emission start of the oscillator A as the phase difference.
Similarly, let $\phi$ be the phase difference with the oscillator C.
From this, the phase difference is derived by

\begin{equation}
	\theta =\frac{Y}{T}*2\pi, \quad \phi = \frac{R}{T}*2\pi.
\end{equation}
These show the synchronization by drawing the phase plane of the phase difference $\theta$ and $\phi$. 

\begin{figure}[h]
	\begin{center}
		\includegraphics[width=8.5cm]{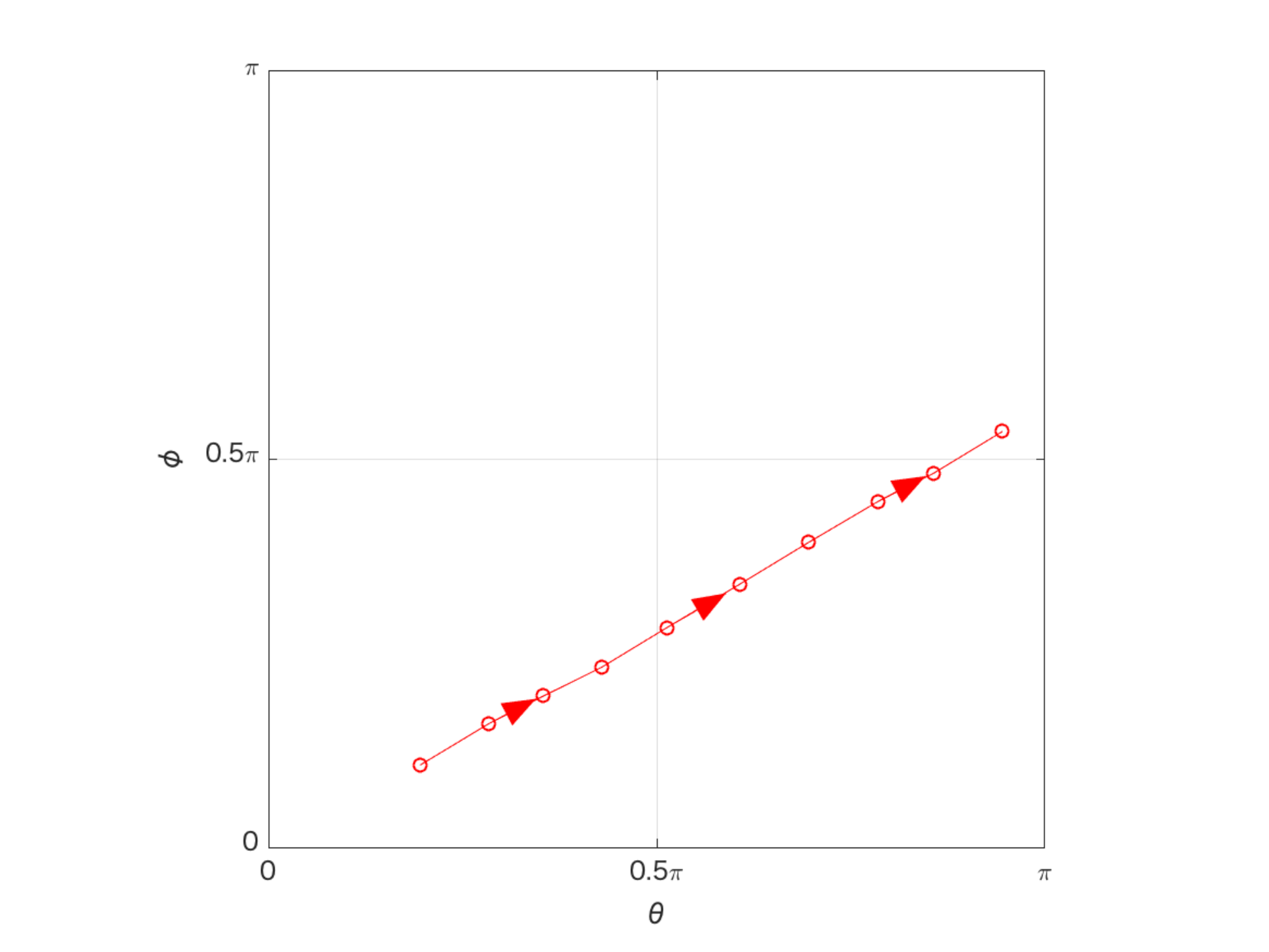}
		\caption{The phase difference of natural vibrations}
		\label{koyu}
	\end{center}
\end{figure}

We now show the experimental results.
First, Fig.\ref{koyu} shows the phase difference of automatically.

The horizontal axis shows the phase difference $\theta$, 
and the vertical axis shows the phase difference $\phi$.

Fig.\ref{koyu} shows a linear function graph that anyone can predict.
Here, we give some notes on how to read graphs.
In this experiment, all three oscillators always light up when the power is turned on.
Therefore, the phase differences $\theta$ and $\phi$ are always 0 at the start of the experiment.
However, since we wanted to show a graph of a linear function in Fig.\ref{koyu} that shows the change in the phase difference of vibration, we intentionally removed the first few analysis results affected by the electromotive force.
From here, two characteristic graphs are shown, showing the process until the three oscillators synchronize and settle into a stable point.

\begin{figure}[h]
\begin{tabular}{cc}
\begin{minipage}{7.5cm}
\begin{center}
\includegraphics[width=8.5cm]{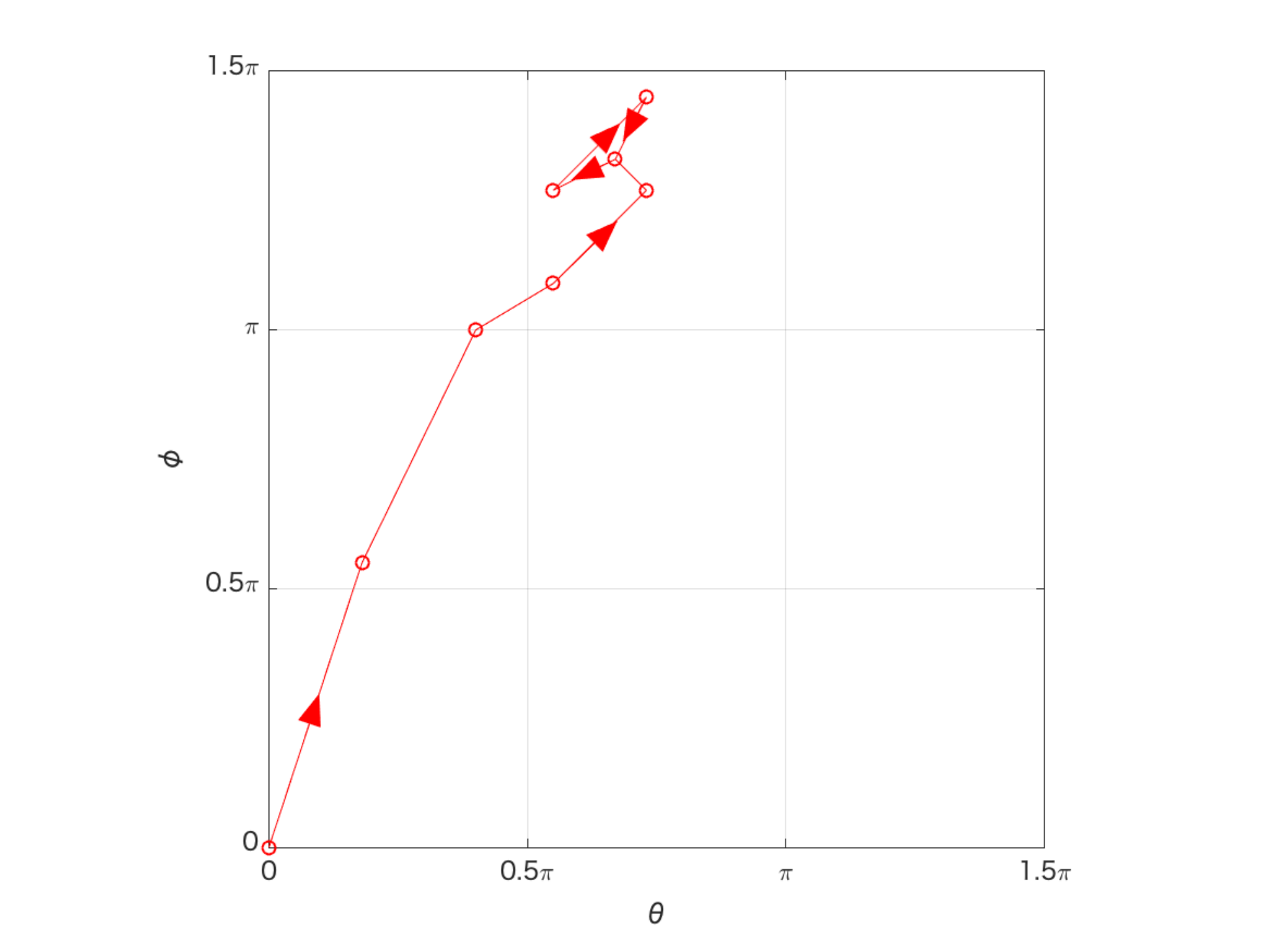}
\caption{A:12cm, B:12cm, C:12cm (Regular Triangle)}
\label{fig:A-12}
\end{center}
\end{minipage}
\hspace{0.5cm}
\begin{minipage}{7.5cm}
\begin{center}
\includegraphics[width=8.5cm]{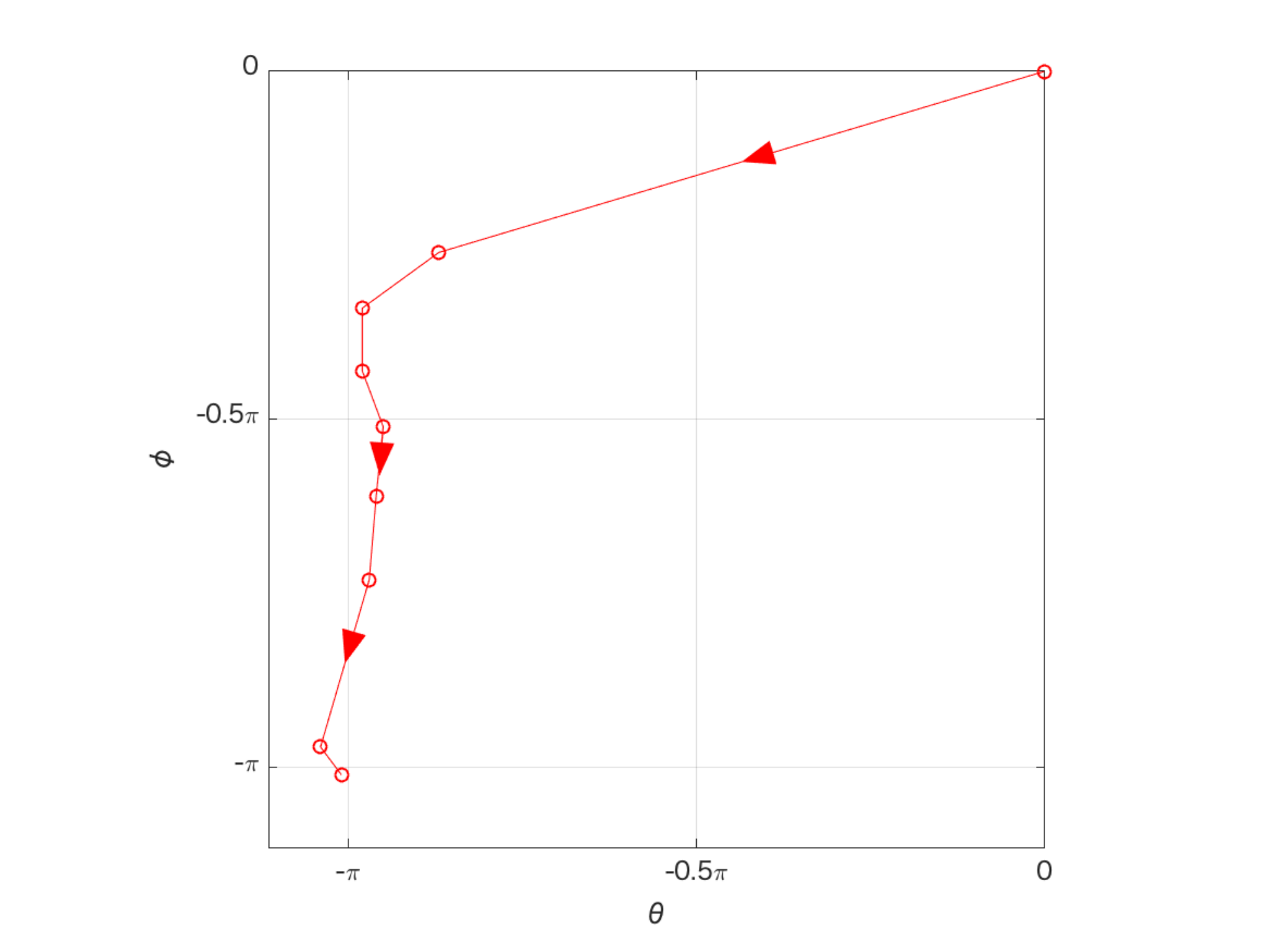}
\caption{A:1cm, B:6cm, C:6cm (Isosceles Triangle)}
\label{fig:A-1}
\end{center}
\end{minipage}
\end{tabular}
\end{figure}

Fig.\ref{fig:A-12} shows the experimental results when three oscillators are placed at a distance of 12cm from the center (Regular Triangle). 
Fig.\ref{fig:A-1} shows the experimental results when the distance from the center is 1cm for one oscillator and 6cm for the other two oscillators (Isosceles Triangle).
In the next section, we show this graph using a mathematical model.

\section{Discussion}

We introduce the well-known "Kuramoto model" \cite{kuramoto1975, Chandra2019, Daido2018, Barre2016} to describe the synchronization phenomenon.
Since the synchronization in this experiment is antiphase synchronization of three oscillators, let $U$ be the total energy of the three oscillators, it can be expressed by
\begin{equation}
	U = \omega_1\theta+\omega_2\phi+\omega_3(\phi-\theta)+\lambda_1\cos(\theta)+ \lambda_2\cos(\phi)+\lambda_3\cos(\phi - \theta).
\end{equation}
Here, $\omega_n(n=1,2,3)$ are natural for frequency, $\lambda_n(n=1,2,3)$ are light coupling strength.
In this experiment, since the circuit design uses the same circuit, it is considered that the difference in natural frequency does not have a great affect, and the term with $\omega_n$ as a coefficient is omitted.
And the phase difference between only two oscillators is considered by using one oscillator as a reference. 
Because of this, we omit the terms whose coefficient is $\omega_n$.

Again, the mathematical model in this experiment is expressed by
\begin{equation}\label{model}
	U = \lambda_1\cos(\theta)+ \lambda_2\cos(\phi)+\lambda_3\cos(\phi - \theta).	
\end{equation}
This equation is focused only on the coupling strength $\lambda_n$ of the oscillator.
From this model, we draw the phase planes of the phase difference $\theta$ and $\phi$.

\begin{figure}[h]
\begin{tabular}{cc}
\begin{minipage}[c]{7.5cm}
\begin{center}
\includegraphics[width=8.5cm]{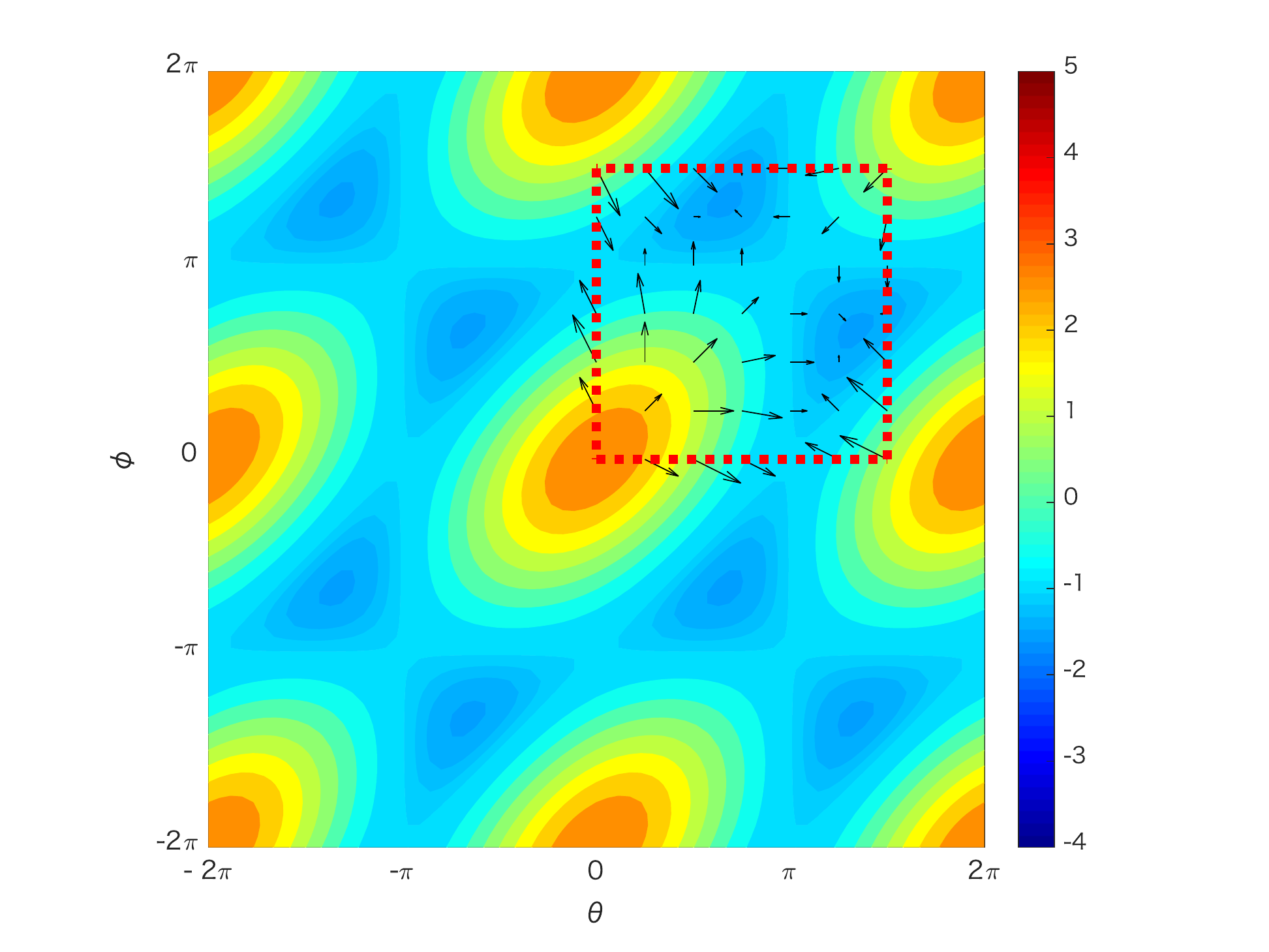}
\caption{$\lambda_{1,2,3}$=1  (The distribution of energy $U$ and Vector)}
\label{fig:spv}
\end{center}
\end{minipage}
\hspace{0.5cm}
\begin{minipage}[c]{7.5cm}
\begin{center}
\includegraphics[width=8.5cm]{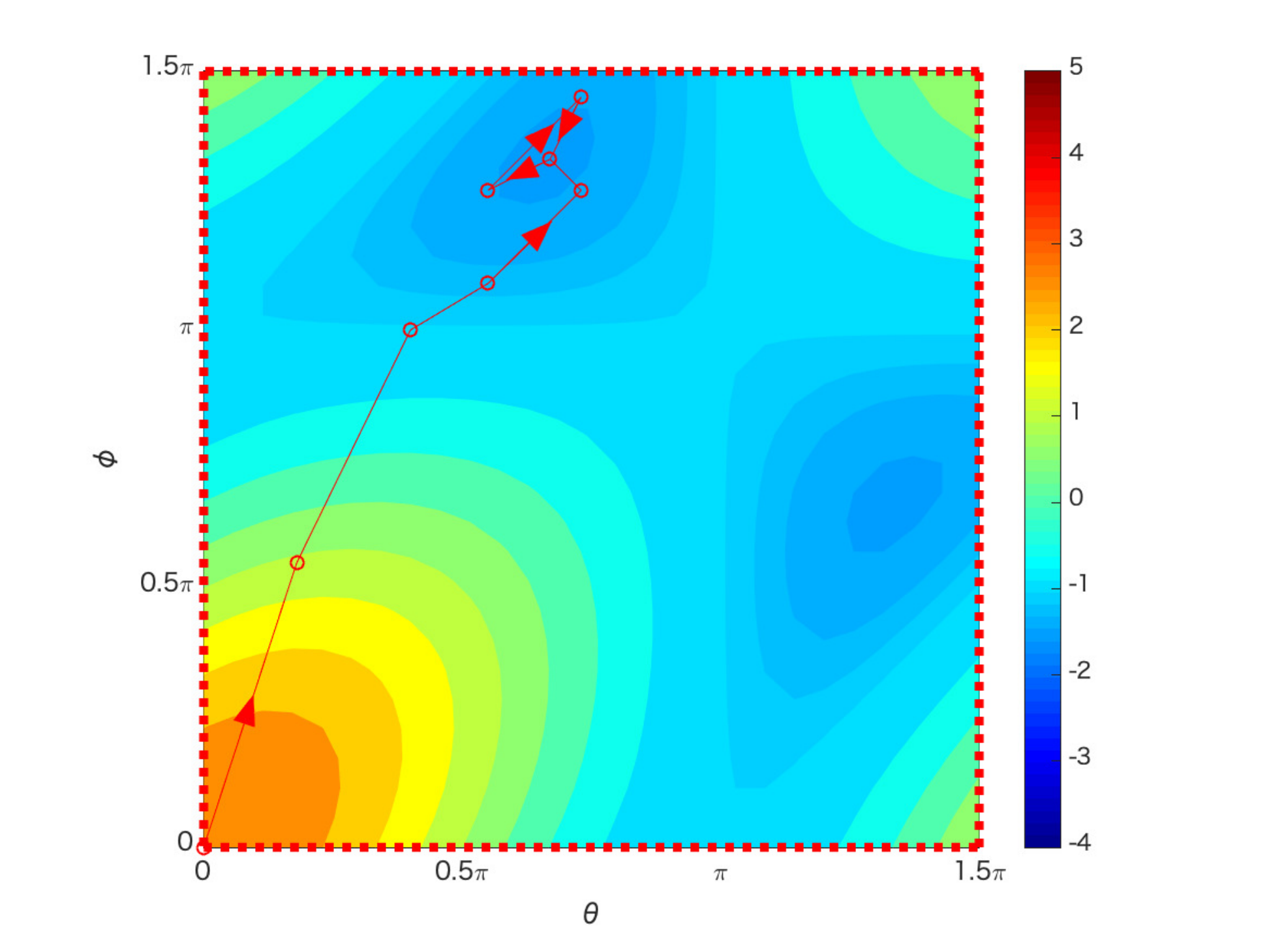}
\caption{$\lambda_{1,2,3}$=1 (The distribution of energy $U$ and Fig.5)}
\label{fig:spp}
\end{center}
\end{minipage}
\end{tabular}
\end{figure}

In the case of Fig.\ref{fig:A-12}, the strength of the coupling of the three oscillators is considered to be equal.
Therefore, Fig.\ref{fig:spv} shows the phase plane when $\lambda_{1,2,3}=1$ in the equation (\ref{model}).

Fig.\ref{fig:spv} shows the distribution of the vector field and the energy $U$.
Fig.\ref{fig:spp} shows the distribution of energy $U$ superimposed on the graph of Fig.\ref{fig:A-12}.
From Fig.\ref{fig:spp}, it can be seen that the experimental result shown in Fig.\ref{fig:A-12} is stable at the lowest energy $U$.

\begin{figure}[h]
\begin{tabular}{cc}
\begin{minipage}{7.5cm}
\begin{center}
\includegraphics[width=8.5cm]{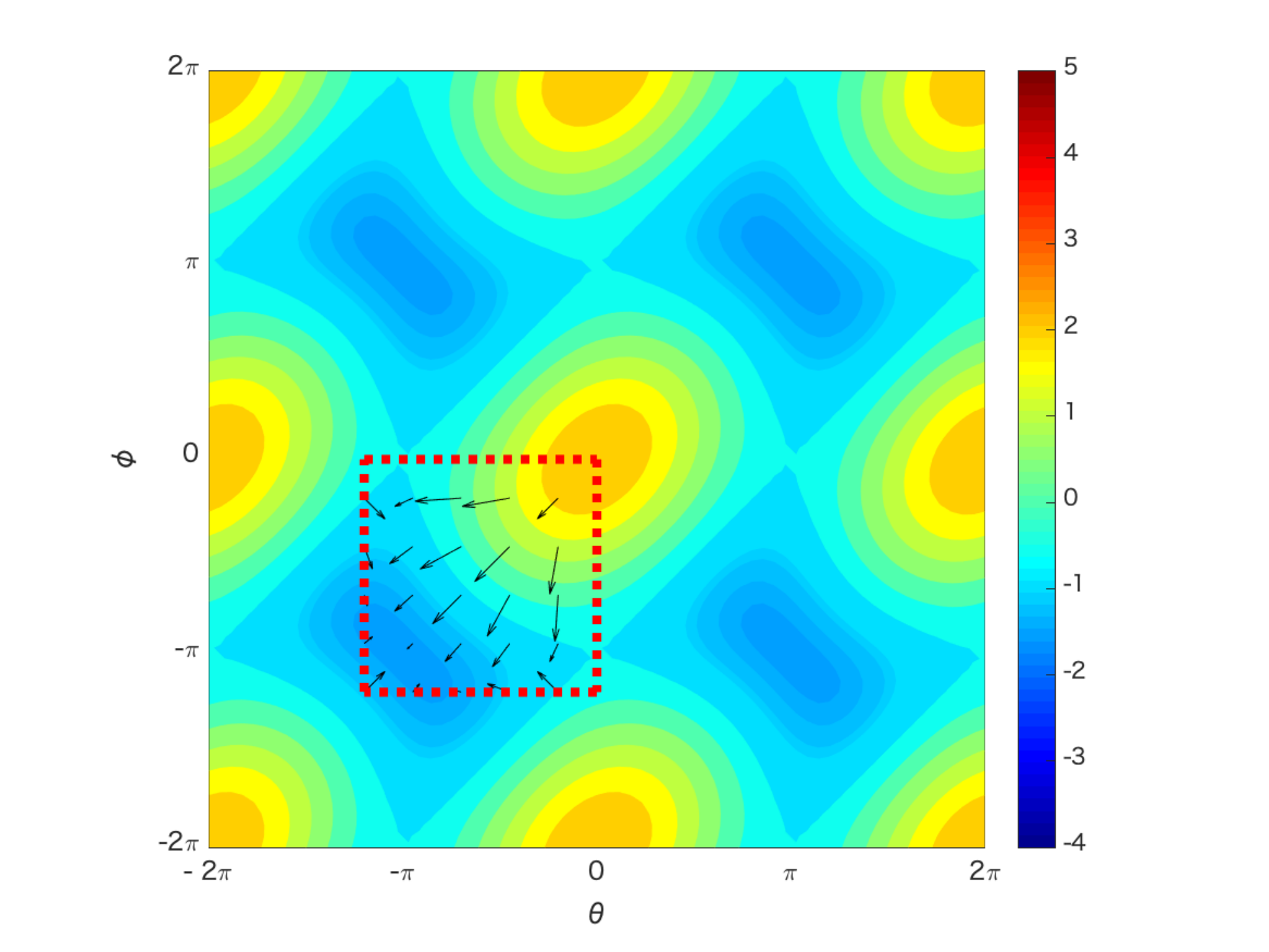}
\caption{$\lambda_{1,2}=1, \lambda_3=0.5$ (The distribution of energy $U$ and Vector)}
\label{fig:owv}
\end{center}
\end{minipage}
\hspace{0.5cm}
\begin{minipage}{7.5cm}
\begin{center}
\includegraphics[width=8.5cm]{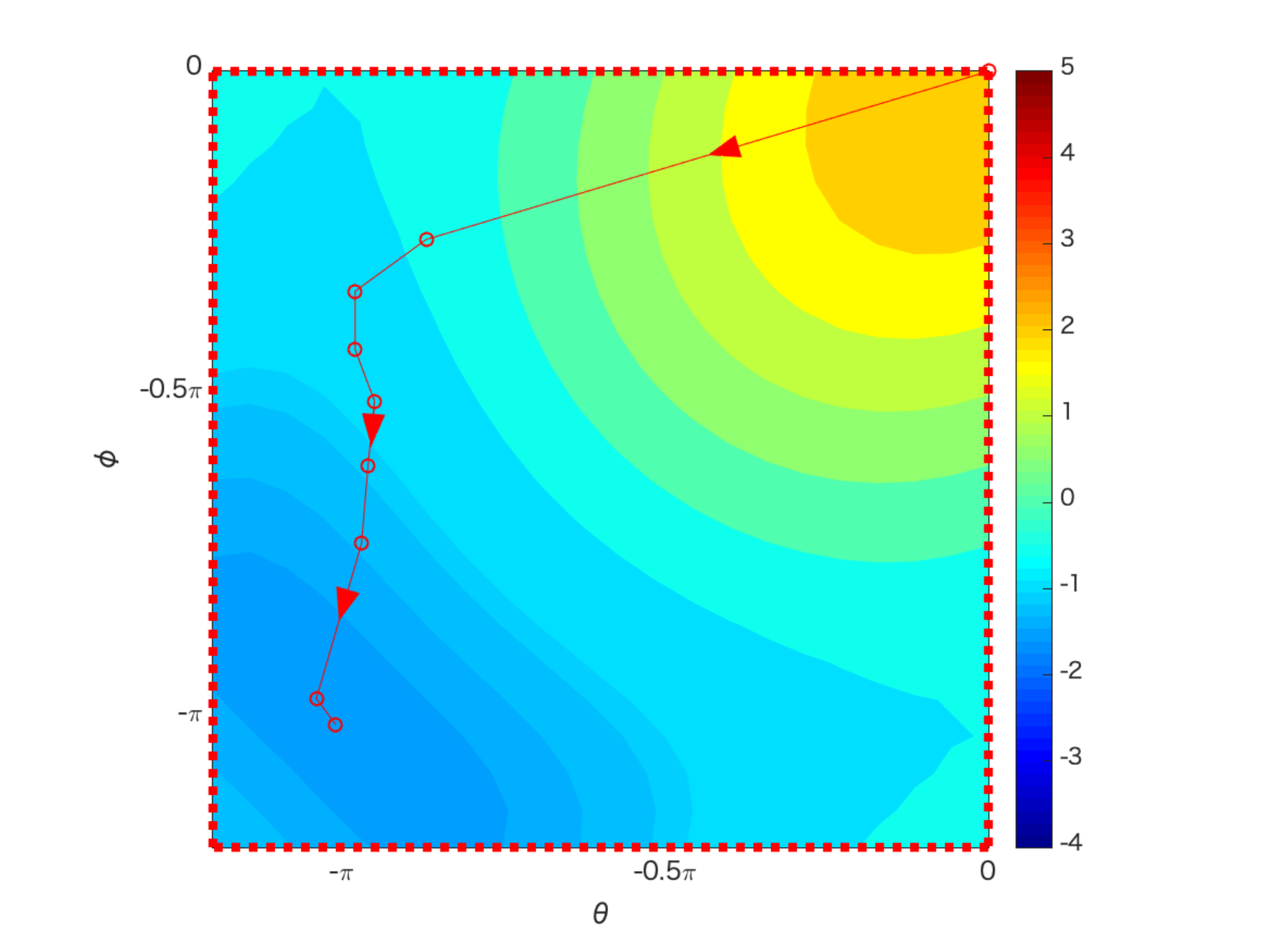}
\caption{$\lambda_{1,2}=1, \lambda_3=0.5$ (The distribution of energy $U$ and Fig.6)}
\label{fig:owp}
\end{center}
\end{minipage}
\end{tabular}
\end{figure}

Next, in the case of Fig.\ref{fig:A-1}, the strength of the coupling of the three oscillators is considered to be weak by only one.
Therefore, Fig.\ref{fig:owv} shows the phase plane when $\lambda_{1,2}=1, \lambda_3=0.5$ in the equation (\ref{model}).

In Fig.\ref{fig:owv}, the vector field and the distribution of energy $U$ are superimposed in the same way as in Fig.\ref{fig:spv}, and in Fig.\ref{fig:owp}, the graph of Fig.\ref{fig:A-1} is superimposed with the distribution of energy $U$.
Even in Fig.\ref{fig:owp}, the experimental result shown in Fig.\ref{fig:A-1} can be seen to be stable at the lowest energy $U$.

From this experiment, it was shown that even if the oscillators are the same, it is possible to create different stable states simply by changing the bond strength.
This can be said to be a network whose behavior changes depending on the initial input.
This could be suggested as a the possibile imitation of brain cell network, as it can be implied for brain cell communication.

In addition, it was possible to show the state of the synchronization state caused by the strength of coupling by a simple mathematical model.
This means that, in other words, you can control the state of synchronization.
However, there are many problems.
The oscillator used this time is the same circuit design.
In theory, such identical synchronization phenomena do not occur if they are exactly the same. 
However, in practice, there is an error in the parts used in the circuit, and a slight difference due to the error shows such multistability.
Therefore, in the experimental system currently in use, noise is inevitably included in the strength of the bond.
For example, subtle differences in resistance depending on elements, performance of LEDs and photo transistors, and slight angular deviations in experimental array.
It is great to be able to see the various syncs, but to control them, you need to overcome them.
Furthermore, this circuit is assumed to be basically the same.
In future, we will try to pursue various synchronization phenomena that can be observed by making a large frequency difference automatically for each oscillator, operating with a time delay, and changing the circuit design.

\end{document}